\documentclass[12pt]{iopart}

\usepackage{iopams,graphicx}  
\begin{document}

\title[Classical particle exchange]{Emergence of long-range force laws from classical particle exchange}

\author{Jarrett L. Lancaster}
\ead{jlancas2@highpoint.edu} %optional optional
\address{Department of Physics, High Point University, One University Parkway, High Point, NC 27262, USA} %

\author{Colin McGuire}
\ead{czmcguire@gmail.com}
\address{The Kiski School, 1888 Brett Lane, Saltsburg, PA 15681, USA}

\author{Aaron P. Titus}
\ead{atitus@highpoint.edu}
\address{Department of Physics, High Point University, One University Parkway, High Point, NC 27262, USA}

%\author{Content \& Services Team}
%
%\address{IOP Publishing, Temple Circus, Temple Way, Bristol BS1 6HG, UK}
%\ead{submissions@iop.org}
\vspace{10pt}
\begin{indented}
\item[]October 2019
\end{indented}

\begin{abstract}
The analogy of classical repulsive interactions emerging from the exchange of mediating particles is revisited with a quantitative approach. Simulations are presented for a particular toy model which are accessible to undergraduate students at any level in the physics curriculum.  Analytic treatment of the various regimes shows rigorously how effective force laws can emerge from an underlying microscopic model and should be accessible to advanced undergraduate physics majors. The analysis presented uses the concept of emergence as motivation for students to gain experience building and testing simplified models for complex physical processes.\end{abstract}

%%%%%%%%%%%
%\begin{document}
%\title{Classical particle exchange: a toy model for action-at-a-distance interactions}
%\author{Jarrett L. Lancaster}
%\email[Electronic mail: ]{jlancas2@highpoint.edu} %optional optional
%\affiliation{Department of Physics, High Point University, One University Parkway, High Point, NC 27262} %

%\author{Colin McGuire}
%\email[Electronic mail: ]{czmcguire@gmail.com}
%\affiliation{Lowcountry Preparatory School, 300 Blue Stem Drive, Pawleys Island, SC 29585}

%\author{Aaron P. Titus}
%\email[Electronic mail: ]{atitus@highpoint.edu}
%\affiliation{Department of Physics, High Point University, One University Parkway, High Point, NC 27262}

%optional
%\date{\today}

%\begin{abstract}

%\end{abstract}
\vspace{2pc}
\noindent{\it Keywords}: emergence, classical field theory, classical particle exchange, force laws, physical models 
%\submitto{\EJP}
%\maketitle

%\section{Introduction: file preparation and submission}

\section{Introduction}

Countless students in introductory physics learn that the ``exchange of virtual particles'' is responsible for the fundamental forces of nature. Several popular introductory textbooks contain diagrams which sketch how classical particle exchange could plausibly explain the qualitative nature of repulsive forces~\cite{BauerWestfall,Mazur}. Furthermore, some texts even attempt to construct analogies for how attractive forces could arise from complicated exchanges of classical objects~\cite{Giancoli,YoungFreedman}. In this paper, we wish to address how such pictures may be {\it quantitatively} useful in understanding the connection between fundamental interactions and momentum transfer through mediating particles. A classical particle exchange model can be used as an interesting problem for students to investigate and also as a bridge toward exploring more advanced, physically correct models for force laws involving classical or quantum fields~\cite{Hilborn}. Such experiences with model development, investigation, exploration of limitations, and falsification are essential for a student learning to ``think like a physicist''~\cite{Etkina}. 

To analyze fundamental interactions properly, the methods of quantum field theory provide the tools necessary for obtaining quantitatively accurate results. Ref.~\cite{Zee} provides a particularly illuminating discussion of how gravitational, electrostatic and nuclear potentials arise as either attractive or repulsive interactions by using the path integral formulation of quantum field theory. Additionally, by casually invoking the energy-time version of the Heisenberg uncertainty principle, one may obtain surprisingly accurate information regarding the force laws resulting from electromagnetic and nuclear interactions~\cite{Harney}, though dimensional analysis plays a large role in constraining the form of the various force laws. The aim of the present work is to explore how effective forces between particles which are spatially separated {\it can} arise within classical particle dynamics. Ultimately, fields are necessary to understand action-at-a-distance interactions fully. However, the jump from mechanics to electromagnetism often lacks a fully convincing motivation for {\it why} fields are a necessary framework for describing such interactions.

In particular, we consider a system of two particles which interact with each other via the exchange of two small, mediating particles. The mediating particles interact with the heavier particles through inelastic collisions, always emerging with speed $c\,$ relative to a stationary lab frame. We explore whether such a system--in which kinetic energy grows in time--could consistently describe interactions between spatially-separated particles by supposing that the increase in kinetic energy can be attributed to a some hidden form of stored, potential energy which is released as the particles gain speed. We show that an effective potential energy for this system can be defined formally. However, this potential energy appears somewhat unusual, and we sketch how a picture in terms of fields can be used to obtain a number of different types of force laws under various assumptions of the model. Though our toy model is quite artificial compared to the quantum field theories describing the known fundamental interactions, the reasoning required for a quantitative analysis is quite useful in understanding the realistic interactions that do occur in nature through mediating quantum fields~\cite{Zee}.  

A notable shortcoming of the classical particle-exchange analogy is its inability to describe attractive forces~\cite{GriffithsPart}. While it is possible to invoke quantum fluctuations in energy to explain attractive nuclear forces in a qualitative manner~\cite{Dunne} we emphasize that attractive interactions emerge naturally from classical scalar field theory~\cite{Rubakov}. Consequently, such treatment is beyond the scope of the present work, as we wish to investigate a model which consists only of classical point particles. Addressing the shortcomings associated with such a model leads naturally to the conclusion that classical fields should replace the mediating particles for internal consistency.

This paper is arranged as follows: in Section~\ref{sec:model} we present a model for classical particle exchange and explore some basic consequences through simulations and physical reasoning, both of which are appropriate for students in introductory physics courses. Section~\ref{sec:analytic} contains a thorough analysis of the model employing advanced physical reasoning and special functions to verify the speculative results obtained through careful estimation in Section~\ref{sec:model}. Finally, we summarize the results in Section~\ref{sec:summary}.

\section{Model}\label{sec:model}

\begin{figure}[h]
\begin{center}
	\includegraphics[totalheight=5.5cm]{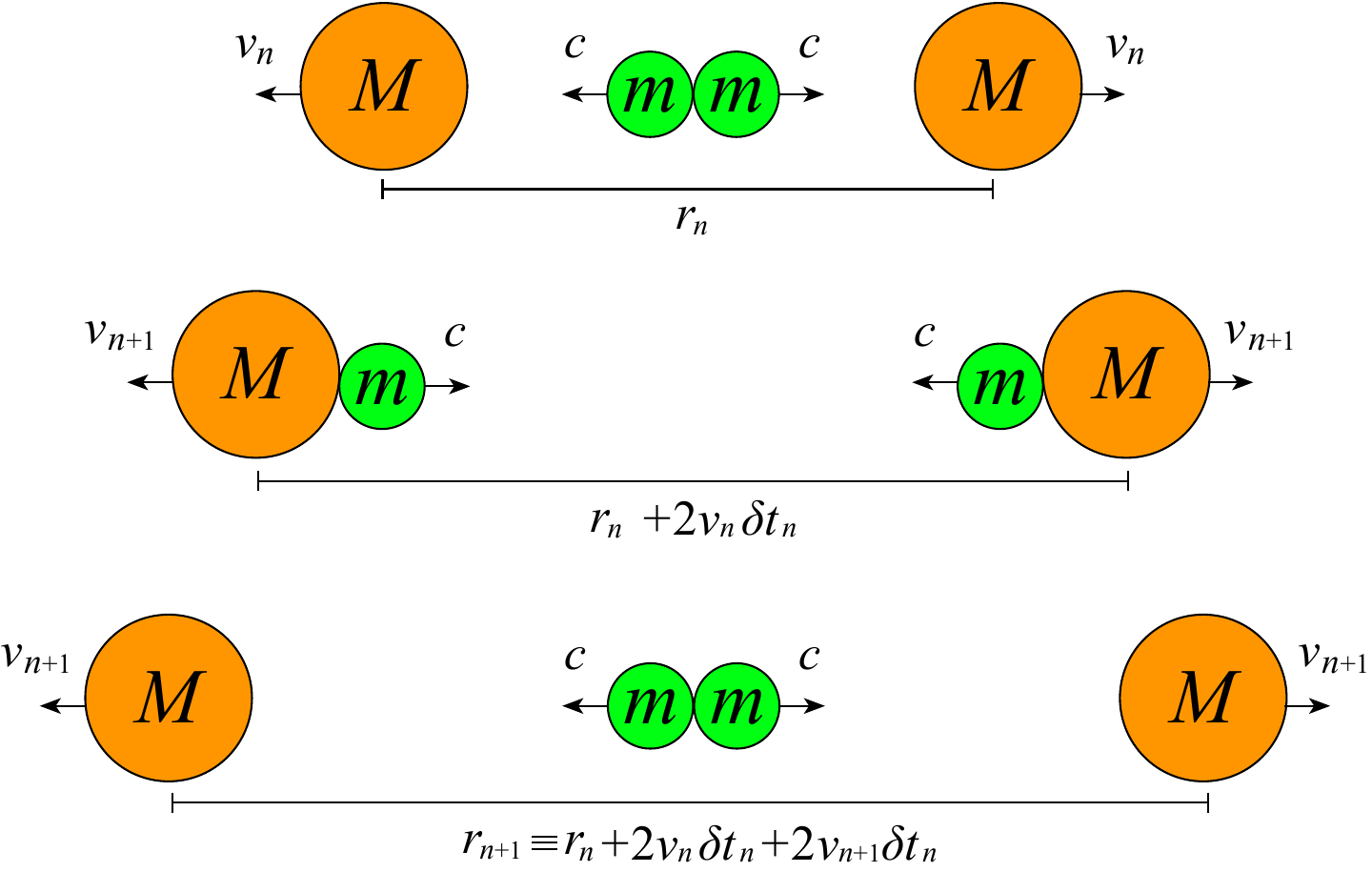}
\caption{Two particles of mass $M\,$ experience a repulsive ``force'' which is mediated by the exchange of a smaller particle of mass $m\ll M$.}
\label{fig:collisionfig}
\end{center}
\end{figure}

Suppose we have two particles each of mass $M\,$ exchanging small particles, each of mass $m\ll M$ as shown in figure~\ref{fig:collisionfig}. The analogy is often made to a pair of ice skaters (or rollerbladers) tossing a ball back and forth~\cite{BauerWestfall,Mazur,Giancoli,YoungFreedman}. Each time one skater catches the ball and throws it back, a small amount of momentum is imparted to that skater, resulting in an effective repulsive force between the skaters which is mediated by the ball being tossed. We construct a quantitative model for this type of interaction by taking each of the smaller particles' speeds to be a constant $c$. The label $c\,$ carries no reference to the actual speed of light, though we will see that our $c\,$ plays a role in our model which is rather similar to that of the actual speed of light in electromagnetism, allowing us to explore a sort of ``non-relativistic'' limit of the model for speeds $v \ll c$.  In order to keep the system's center of mass at rest, we consider a symmetric setup in which two small particles are exchanged. When the smaller, mediating particles approach each other we assume that they collide elastically.

Imposing the rule that the mediating particles always move at speed $c$ means that their kinetic energy remains constant. However, with each collision of a mediating particle and larger particle, the larger particle's momentum magnitude and kinetic energy increase, making the collisions inelastic. We complete the specification of the model by postulating some hidden degrees of freedom are responsible for the maintaining the speed of the mediating particles, thereby allowing the total kinetic energy of the mechanical system to increase. We demonstrate below that the work required for the observed changes in kinetic energy may be associated formally with an effective ``potential energy'' for the system which decreases as the particles gain speed. Taking the large, right-moving particle to be moving at speed $v_{n}$ ($v_{n+1}$) before (after) the collision, momentum conservation applied to a single collision gives
\begin{equation}
Mv_{n} + mc = Mv_{n+1} - mc,
\label{Eq:momentumconservation}
\end{equation}
or $\delta v \equiv v_{n+1}-v_{n} = 2\frac{mc}{M}$. With repeated collisions of this form, the two massive particles will accelerate away from their common center of mass in a manner qualitatively similar to the motion experienced by two like charges placed near each other and released. We employ two approaches to investigate the quantitative nature of this effective force law. First, we simulate the system as described, obtaining numerically an effective force law which decreases as $r^{-1}\,$ for small velocities $v\ll c$, where $r\,$ is the instantaneous separation between the two massive particles. Second, the discrete sequence of collisions leads to a recursion relation which allows us to obtain a closed-form expression for $r_{n}$, the separation distance immediately preceding the $n^{\mbox{\scriptsize th}}\,$ collision. In the limit of small velocities, $v\ll c$, both approaches allow us to extract an effective long-range force which acts on the large particles as a power-law function of separation distance. 
\subsection{Full simulation}
\begin{figure}[h]
\begin{center}
	\includegraphics[totalheight=7.2cm]{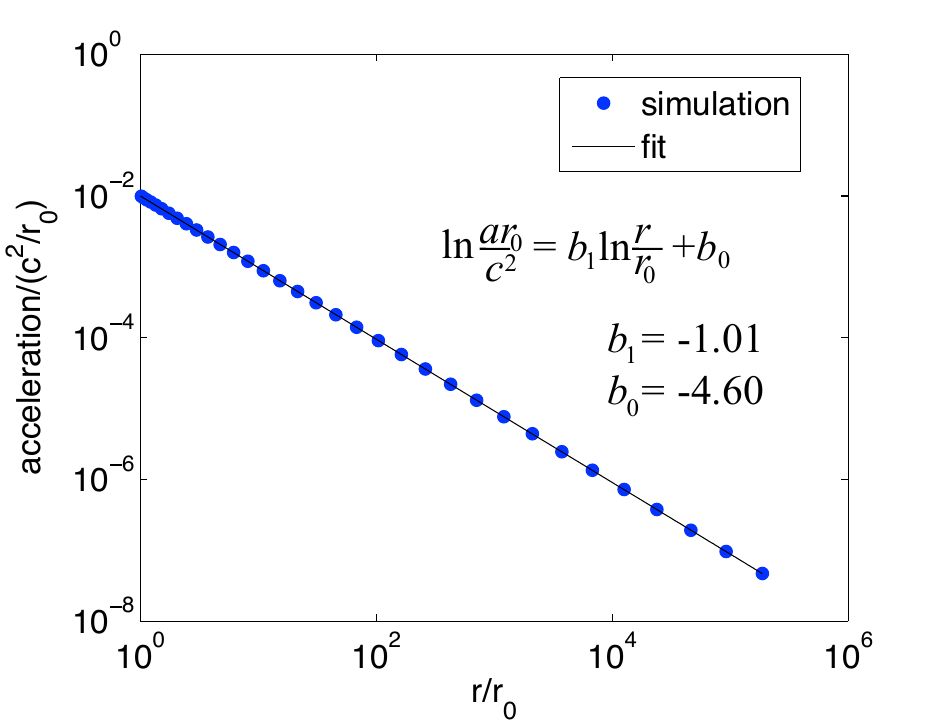}
\caption{Numerically computed acceleration is plotted against separation distance for the right-moving mass with logarithmic scales on axes. Also shown is a linear regression for the logarithmic data.}
\label{fig:loglog}
\end{center}
\end{figure}

The full simulation consists of integrating the Newtonian equations of motion for free particles moving at constant speeds and monitoring for a ``collision'' at which point each massive particle is given a boost in speed $\delta v = 2mc/M\,$ and the mediating particles are reflected with equal momenta in the opposite directions. Letting $x^{(1)}\,$ ($x^{(2)}$) denote the position of the right-moving (left-moving) particle and $v^{(1)}\,$ ($v^{(2)}$) its velocity, we take the initial separation between the particles to be $r_{0}$. The corresponding initial conditions are
\begin{eqnarray}
x^{(1)}(0) & = & -x^{(2)}(0) = \frac{r_{0}}{2},\\
v^{(1)}(0) & = & v^{(2)}(0) = 0.
\end{eqnarray}
The mediating particles are initially located at the origin and begin moving in opposite directions toward the massive particles at $t = 0$, each with speed $c$. Letting the positions of the mediating particles be given by $X^{(i)}\,$ for $i=1,2$, it is an instructive exercise to numerically integrate the equations of motion
\begin{eqnarray}
\frac{dx^{(i)}}{dt} & = & v^{(i)},\\
\frac{dX^{(i)}}{dt} & = & V^{(i)},
\end{eqnarray}
with $V^{(1)} = +c\,$ and $V^{(2)} = -c\,$ at $t=0$. To monitor for collisions, at each time step $\Delta t\,$ we check for the following condition:
\begin{equation}
\left| x^{(i)} - X^{(i)} \right| < \epsilon,
\end{equation}
indicating that the mediating particle nearest the $i^{\mbox{\scriptsize th}}\,$ particle has come within a small distance $\epsilon\,$ of the massive particle's location. When this occurs, we make the following adjustment to the equations of motion:
\begin{eqnarray}
v^{(i)} & \rightarrow & v^{(i)} + \frac{2mc}{M}\mbox{sign}\left(V^{(j)}\right),\\
V^{(i)} & \rightarrow & - V^{(i)},
\end{eqnarray}
indicating that a collision has occurred, resulting in momentum transfer. Results are generally insensitive to the time-step size, provided $\epsilon \precsim c\Delta t$. Figure~\ref{fig:loglog} depicts the numerically computed average acceleration as a function of separation distance for $m = 0.005M$. For the computation of acceleration, we use the separation distance and corresponding time just after collision events, since each massive particle's acceleration is formally zero between collisions. Note that for position measurements which are taken at unequal time increments, we require the following discrete representation~\cite{Abramowitz} of the second derivative
\begin{equation}
\left.\frac{d^{2}r}{dt^{2}}\right|_{r=r_{n}} \approx \frac{\frac{r_{n+1}-r_{n}}{t_{n+1}-t_{n}}-\frac{r_{n}-r_{n-1}}{t_{n}-t_{n-1}} }{t_{n+1}-t_{n-1}}.
\end{equation}

A strong linear trend on a log-log plot demonstrates the power-law nature of the force law,
\begin{equation}
\frac{d^{2}x^{(2)}}{dt^{2}} \propto r^{b_{1}},
\end{equation}
with $b_{1} \approx -1$. This result is consistent with a rough estimation of the rate of momentum transfer for $v \ll c$. Each collision is associated with transfer of momentum 
\begin{equation}
\delta p = 2mc.
\end{equation}
For $v \ll c$, the massive particles do not move appreciably during one collision cycle. Let $r\,$ denote the instantaneous separation distance between the massive particles. Beginning with the mediating particles at the origin, one cycle requires each particle to cover a distance $\frac{r}{2}\,$ to collide with the massive particles and then another distance of $\frac{r}{2}\,$ to return to the origin. Thus, a single collision cycle associated with a momentum transfer $\delta p\,$ requires a time
\begin{equation}
\delta t \simeq \frac{r}{c}.
\end{equation}
The average force experienced by each massive particle is then
\begin{equation}
F_{\mbox{\scriptsize ave}} \simeq \frac{\delta p}{\delta t} = \frac{2mc^{2}}{r}.\label{eq:force}
\end{equation}
The validity of this crude estimate will be examined more carefully in the next section, but for now it serves to make the results in figure~\ref{fig:loglog} appear rather plausible. Here, the result depends on $v \ll c$ so that the massive particles do not move appreciably during a collision cycle. To examine the validity of~(\ref{eq:force}) more systematically and probe the high-velocity limit of the model, we must refine the simulation method in order to access much longer times. 

\subsection{Calculation of collision times}
The results so far suggest a disconnect between the low-energy behavior of the model and the high-energy ``speed limit'' of $c$, which should be enforced by the mediating particles. To obtain some understanding of the large-speed regime, we must explore extremely large timescales, thus allowing the larger particles to approach high speeds, $v^{(1,2)}\sim c$. Because the time between subsequent collisions grows at an accelerated rate as the massive particles spread apart and speed up, a brute-force numerical integration of the equations of motion to study the time evolution of the system becomes impractical. In fact, most of the computation is fundamentally unnecessary since all particles move with constant velocities until a collision occurs. Starting from one collision event, the time that elapses before the next collision may be computed using the instantaneous velocities of all particles, and this process may be repeated. Though the time between collisions grows rapidly, the computation time of this scheme grows linearly with number of collisions rather than the elapsed time as for a direct integration of the equations of motion. 

To proceed, let us consider a single collision event shown in figure~\ref{fig:collisionfig}. With both mediating particles located at the origin and instantaneous separation $r_{n}\,$ between the outwardly moving massive particles, the next collision will occur after the mediating particles have reached the massive particles, requiring a time
\begin{equation}
\delta t_{n} = \frac{r_{n}/2}{c-v_{n}},\label{eq:dt}
\end{equation}
corresponding to traveling a distance of $\frac{r_{n}}{2}\,$ with speed $c-v_{n}\,$ relative to the outwardly moving, massive particles. After time $\delta t_{n}\,$ has elapsed, collisions occur resulting in the mediating particles reversing directions and
\begin{equation}
v_{n}\rightarrow v_{n+1} \equiv v_{n} + \frac{2mc}{M}.\label{eq:vn}
\end{equation}
The cycle completes when the mediating particles return to the origin. By symmetry, this return time also requires time $\delta t_{n}$, so the entire elapsed time for a complete cycle is $2\delta t_{n}$, or
\begin{equation}
t_{n+1} = t_{n} + \frac{r_{n}}{c-v_{n}}.\label{eq:tn}
\end{equation}
To update the positions of the massive particles, we note that before the collision, each particle was moving away from the center of mass with speed $v_{n}$ with respect to the lab frame for time $\delta t_{n}$. After the collision, each particle moves away from the system's center of mass for time $\delta t_{n}\,$ with the updated speed, $v_{n+1}$. Thus, the separation distance increases by an amount $2v_{n}\delta t_{n} + 2v_{n+1}\delta t_{n}$, or
\begin{equation}
r_{n+1} = r_{n} + 2v_{n}\delta t_{n} + 2v_{n+1} \delta t_{n}.\label{eq:rn}
\end{equation}
Equations~(\ref{eq:dt})--(\ref{eq:rn}) constitute a closed recursion relation which may be iteratively advanced to obtain the velocity, separation distance and time corresponding to the beginning of each collision cycle. 

We wish to compare the exact results in (\ref{eq:dt})--(\ref{eq:rn}) to the approximate, low-speed predictions which arise from a force law of the form (\ref{eq:force}). An approximate, analytic solution in the regime where (\ref{eq:force}) applies may be obtained by writing Newton's second law for the motion of the right-moving particle,
\begin{equation}
M\frac{d^{2}x^{(1)}}{dt^{2}} = \frac{2mc^{2}}{r}.
\end{equation}
Applying the symmetry of the system, we have $r = 2x^{(1)}\,$ and may change variables,
\begin{equation}
\frac{d^{2}x^{(1)}}{dt^{2}} = \frac{d^{2}r}{dt^{2}} = \frac{1}{2}\frac{d}{dr}\left(\dot{r}^{2}\right).
\end{equation}
Writing $\dot{r} = 2v$, where $v\,$ represents the speed of each massive particle, we may integrate both sides to obtain
\begin{equation}
v^{2}(r) = v_{0}^{2} + \frac{2mc^{2}}{M}\ln\frac{r}{r_{0}}.\label{eq:nrap}
\end{equation}
\begin{figure}[h]
\begin{center}
	\includegraphics[totalheight=7.2cm]{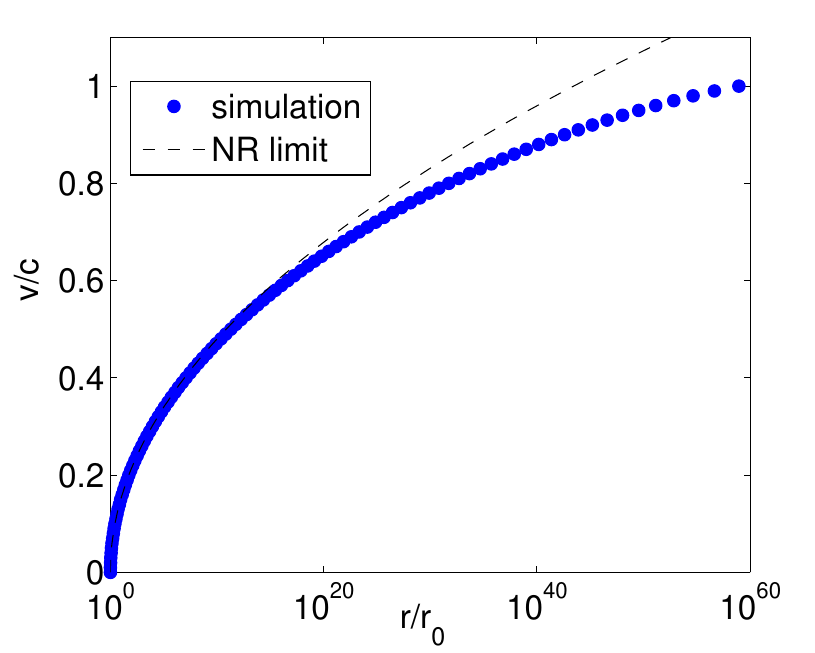}
\caption{Long-time, large-distance behavior of massive particle speed (blue circles) computed from~(\ref{eq:dt})-(\ref{eq:rn}) and compared to low-speed, non-relativistic (NR) approximation in~(\ref{eq:nrap}), which provides excellent agreement with the simulation for $v \ll c$.}
\label{fig:logplot1}
\end{center}
\end{figure}

We refer to~(\ref{eq:nrap}) as the {\it non-relativistic approximation}, as its derivation relies on assuming $v\ll c$. The term ``non-relativistic'' (NR) as used here does not refer to speeds much less than the actual speed of light but those significantly smaller than the mediating particle speed $c$. The role played by $c\,$ in this model is similar to that of the actual speed of light in electrodynamics, but we stress that special relativity and the actual speed of light play no role in this model.

Figure~\ref{fig:logplot1} depicts the predictions of ~(\ref{eq:nrap}) compared to the actual simulation information contained in~(\ref{eq:dt})--(\ref{eq:rn}). As expected, the non-relativistic approximation breaks down as the massive particles' speeds approach $c$. For large separation distances, the massive particle speeds do not increase as sharply with increasing distance as the non-relativistic approximation predicts. Indeed, once the massive particles reach a speed of $c$, the mediating particles, also traveling at speed $c$, are unable to catch up to the massive particles. Correspondingly the recursion relations break down and no more collisions are found. Specifically, as $v_{n}\rightarrow c\,$ from below, we have $\delta t_{n}\rightarrow \infty$. If the massive particle speed becomes exactly $c$, $\delta t_{n}\,$ does not exist and no further collisions occur. Another possibility is that a single collision changes $v_{n}\,$ from just below $c\,$ to just above $c$. In this case, $\delta t_{n}\,$ formally becomes negative and we conclude similarly that no further collisions occur.

Equation~(\ref{eq:nrap}) formally resembles a one-dimensional statement of the work-energy principle with the corresponding potential energy given by 
\begin{equation}
U(r) = \frac{2mc^{2}}{M}\ln \frac{a}{r},
\label{eq:potentialenergy}
\end{equation}
for some arbitrary length scale $a$.  Here we have {\it defined} $U(r)$ as a bookkeeping device to balance the increasing kinetic energy in the mechanical system. Whether such a definition is sound must be examined (see below), but it should be noted that the entire notion of potential energy is little more than a convenient bookkeeping device, as potential energy is not by itself measurable~\cite{Hecht}. It should be noted that the notion of potential energy arises here by restricting attention to the subsystem consisting of only the two massive particles. The potential energy emerges as an effective representation of role played by the mediating particles and whatever hidden degrees of freedom ensures that they maintain their constant speeds after collisions. 

The behavior of the system explored thus far can be summarized as follows: for arbitrary initial separations, the massive particles are repelled from each other by the effective force provided by the mediating particles. At long times, the speeds (with respect to the ground) of the massive particles approach the speed of the mediating particles, $c$. While an approximate statement of energy conservation has been derived (c.f.,~(\ref{eq:nrap})) for low speeds $v \ll c$, the associated potential is problematic as it has no lower bound for $r\rightarrow \infty$. An unlimited amount of potential energy may be converted into the massive particles' kinetic energy resulting in the erroneous prediction that for any initial separation, both massive particles will continue to accelerate rather than asymptotically approach finite speeds. That the initial separation distance has no effect on the final speeds of the massive particles is highly unusual. In the next section, we will carefully examine this system using analytic tools to quantitatively explore some of these issues.

\section{Analytic approach}\label{sec:analytic}

\subsection{Exact solution to recursion relation}
The discrete sequence of collisions described by~(\ref{eq:dt})--(\ref{eq:rn}) can be analyzed exactly, yielding a closed-form expression for $r_{n}$. Equation~(\ref{eq:vn}) simply states that the velocity increases by a constant amount after each collision, or
\begin{equation}
v_{n} = \frac{2mnc}{M}.\label{eq:vnsol}
\end{equation}
Inserting~(\ref{eq:vnsol}) into~(\ref{eq:rn}) and using~(\ref{eq:dt}), we have
\begin{eqnarray}
r_{n+1} & = & r_{n} + 2\left[v_{n} + \frac{mc}{M}\right]\frac{r_{n}}{c-v_{n}},\\
& = & \left(\frac{1 + \frac{2m(n+1)}{M}}{1 - \frac{2mn}{M}}\right)r_{n}.
\end{eqnarray}
Proceeding iteratively,
\begin{eqnarray}
r_{1} & = & \left(1 + \frac{2m}{M}\right)r_{0},\\
r_{2} & = & \frac{\left(1 + \frac{4m}{M}\right)\left(1 + \frac{2m}{M}\right)}{\left(1-\frac{2m}{M}\right)}r_{0},\\
& \vdots & \\
r_{n} & = & \left(1 + \frac{2nm}{M}\right)\prod_{k=0}^{n-1}\left(\frac{1 + \frac{2km}{M}}{1 - \frac{2km}{M}}\right)r_{0}.\label{eq:rn1}
\end{eqnarray}
By employing the gamma function, which satisfies~\cite{ArfkenWeber} $\Gamma(x+1) = x\Gamma(x)$, and reduces to the factorial for integer arguments, $n! = \Gamma(n+1)$, we may write~(\ref{eq:rn1}) as
\begin{equation}
r_{n} = \frac{\Gamma\left(\frac{M}{2m} + n\right)\Gamma\left(\frac{M}{2m}-n\right)}{\left[\Gamma\left(\frac{M}{2m}\right)\right]^{2}}\left(1 - \left(\frac{2mn}{M}\right)^{2}\right)r_{0}.\label{eq:rn2}
\end{equation}
The derivation of~(\ref{eq:rn2}) from~(\ref{eq:rn1}) requires the properties~\cite{ArfkenWeber}
\begin{eqnarray}
\Gamma(x)\Gamma(1-x) & = & \frac{\pi}{\sin(\pi x)},\;\;\;\;\;\;
\Gamma(x)\Gamma(-x) = -\frac{\pi}{x\sin(\pi x)}.
\end{eqnarray}
\subsection{Limiting cases}
As an exact, closed-form solution,~(\ref{eq:rn2}) contains all of the physics we have encountered up to this point. The low-energy force law in~(\ref{eq:force}) was previously derived using physical reasoning, but we can demonstrate that it also follows from the exact solution rather than appealing to comparisons such as~(\ref{fig:loglog}). To this end, let us define $\alpha \equiv \frac{M}{2m}\,$ and take the natural logarithm of~(\ref{eq:rn2}), obtaining
\begin{eqnarray}
\ln \frac{r}{r_{0}} & = & \ln \Gamma\left(\alpha-n\right) + \ln \Gamma \left(\alpha - n\right) - 2\ln\Gamma \left(\alpha\right)+ \ln\left[1 - \left(\frac{n}{\alpha}\right)^{2}\right].
\end{eqnarray}
To investigate the dynamics for $m \ll M\,$ and $v \ll c$, we examine the limit $\alpha \rightarrow \infty\,$ with $n \ll \alpha$. We first apply Stirling's approximation~\cite{ArfkenWeber} to the Gamma functions,
\begin{eqnarray}
\ln \Gamma\left(\alpha \pm n\right) & \simeq & \left(\alpha \pm n\right)\ln\left[\alpha \pm n\right],
\end{eqnarray}
Applying the limit $n \ll \alpha\,$ and expanding the logarithms according to $\left(1\pm x\right)\ln \left[1 \pm x\right] \simeq x + \frac{x^{2}}{2}$, we recover the result
\begin{equation} 
\ln \frac{r}{r_{0}} \simeq \frac{n^{2}}{\alpha},
\end{equation}
which is equivalent to~(\ref{eq:nrap}) with $v_{0} = 0\,$ upon the identification $n\rightarrow \frac{M}{2m}\frac{v}{c}\,$ (c.f.,~(\ref{eq:vnsol})).

Alternatively, we may consider the limit $v\rightarrow c$. Note that~(\ref{eq:rn2}) diverges as $n\rightarrow \alpha$, indicating that this only occurs as $r\rightarrow \infty$. Implicit in this relation is the upper limit on number of collisions before the massive particles reach terminal velocity,
\begin{equation}
n_{\mbox{\scriptsize max}} = \frac{M}{2m}.
\end{equation}
We may probe the system at long times by letting $n = \alpha - \epsilon\,$ for $\epsilon \ll 1$. Equation~(\ref{eq:rn2}) then becomes
\begin{equation}
\frac{r_{n}}{r_{0}} = \frac{\Gamma\left(2\alpha\right)}{\left[\Gamma\left(\alpha\right)\right]^{2}}\Gamma\left(\epsilon\right)\cdot \frac{2\epsilon}{\alpha}.\label{eq:smallv1}
\end{equation}
Employing the small-argument expansion~\cite{PeskinSchroeder}
\begin{equation}
\Gamma\left(\epsilon\right) = \frac{1}{\epsilon} - \gamma +\mathcal{O}(\epsilon),
\end{equation}
where $\gamma \simeq 0.577\,$ is the Euler-Mascheroni constant, we may expand ~(\ref{eq:smallv1}) to obtain
\begin{equation}
\frac{r}{r_{0}} \simeq \frac{2\Gamma\left(2\alpha\right)}{\alpha\left[\Gamma(\alpha)\right]^{2}}\left(1-\gamma\epsilon\right).
\end{equation}
Taking $\epsilon\rightarrow 0\,$ is equivalent to letting $v\rightarrow c$, and in this limit we obtain the critical separation $r_{c}$ at which $v$ approaches $c$,
\begin{equation}
r \rightarrow r_{c}\equiv \frac{4m\Gamma\left(\frac{M}{m}\right)}{M\left[\Gamma\left(\frac{M}{2m}\right)\right]^{2}}r_{0}\;\;\;\;\;\mbox{ as }v\rightarrow c.\label{eq:rc}
\end{equation}
For $r>r_{c}$, each massive particle is moving at the same speed as the mediating particles and experiences no subsequent collisions with the mediating particles. Some mystery may be removed from~(\ref{eq:rc}) by taking the natural logarithm of both sides and applying Stirling's approximation, this time keeping several terms
\begin{equation}
\ln \Gamma (x) \simeq x\ln x - x - \frac{1}{2}\ln \frac{x}{2\pi}.
\end{equation}
When the smoke clears, we have the compact result
\begin{equation}
r_{c} = \sqrt{\frac{2m}{\pi M}}2^{\frac{M}{m}}r_{0}.
\end{equation}
That is, at a finite separation distance, the massive particles attain their maximum speeds $v = c$. While the potential energy function in (\ref{eq:potentialenergy}) appears to describe the dynamics for $v \ll c$, it continues to decrease without bound as $r\rightarrow \infty$. This in itself is not catastrophic, as the actual Coulomb potential breaks down when charges approach speeds comparable to the speed of light~\cite{GriffithsEM}.

Essentially, we have found that despite the suspicious increase in the mechanical system's kinetic energy with time, the low-energy dynamics formally resemble motion under the influence of a conservative force through (\ref{eq:nrap}). The {\it origin} of this force is somewhat mysterious, as it arises by requiring momentum conservation and the constant speed of mediating particles. Though artificial in this example, such a situation arises in electrostatics where the origin of action-at-a-distance forces between charges may be attributed to electric {\it fields} which exist as independent degrees of freedom from the particles. By analogy, we can postulate the existence of some sort of background field in this toy model which, like electric fields, can store energy that can be converted into mechanical energy. This emergence of an energy-storing field concept, from a purely mechanical system, could be useful in motivating the abstract fields necessary to investigate electromagnetism at the introductory level.

\subsection{Scaling arguments}
The main result of this analysis is the prediction of an inverse-linear force law between two particles which exchange mediating particles in such a way that momentum is conserved and the mediating particle speed $c$ remains constant. In this section we explore how the basic features of the particle-exchange model could be modified to adjust the resulting force law. For example, actual forces resulting from classical field (or quantum fields at ``tree level'')~\cite{Zee} are generically inverse-{\it square} laws. It is the purpose of this section to explore which types of general changes could result in our model predicting an inverse-square law. Despite the limitations of dealing with a toy model, some insight into forces and fields can be gained by exploring the relationship between model structure and resulting force law.

For comparison, we note that a similar force law can be obtained by requiring conservation of kinetic energy in addition to conservation of momentum during each collision. In such a model, the force obtained~\cite{Ee} is given by
\begin{eqnarray}
F & = & \frac{2mc_{0}^{2}}{r^{3}},\label{eq:force2}
\end{eqnarray}
where $c_{0}$ is the initial speed of each mediating particle. In order to conserve kinetic energy, the mediating particles slow down during each collision. Defining $m_{\mbox{\scriptsize eff}}\equiv m\left(\frac{r_{0}}{r}\right)^{2}$, the force in~(\ref{eq:force2}) formally looks identical to the effective force obtained in~(\ref{eq:force}), but with an ``effective mass'' that decreases inversely with $r^{2}$,
\begin{equation}
F\rightarrow \frac{2m_{\mbox{\scriptsize eff}}c_{0}^{2}}{r}.
\end{equation}
Interestingly, this {\it would} correspond to an inverse-square law if the additional factor of $r^{-1}\,$ were absent. This additional factor of inverse distance can be traced to the time between collisions, which scales as $\delta t \sim r/c\,$ and appears in the denominator of~(\ref{eq:force}). 

\begin{figure}[h]
\begin{center}
	\includegraphics[totalheight=4cm]{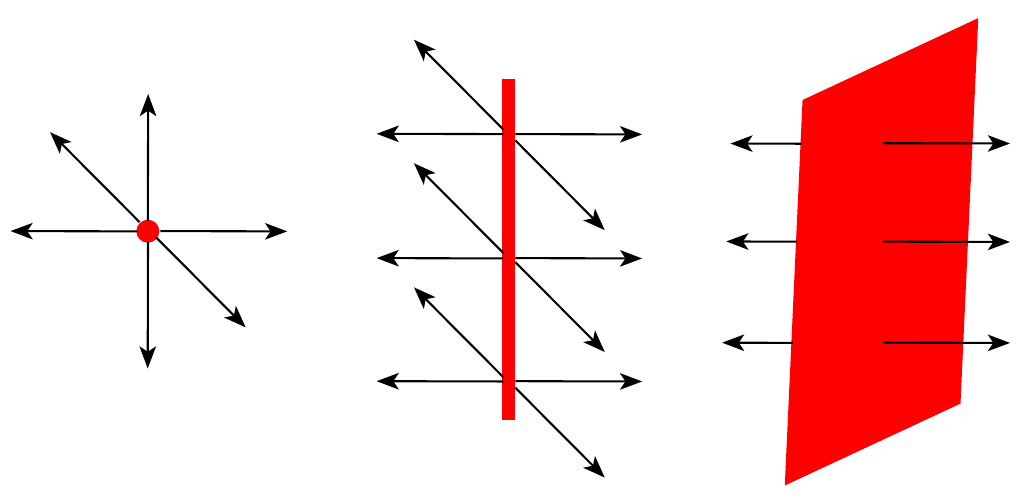}
\caption{A point source (left) gives rise to a field which spreads out in three spatial dimensions and decreases as $E\sim r^{-2}$. For an infinite wire of uniform charge density (center), the field is squeezed into the two remaining dimensions with a slower decay, $E\sim r^{-1}$. In an infinite sheet of uniform charge density (right), only one dimension remains for the field, which is constant $E\sim r^{0}$.}
\label{fig:efields}
\end{center}
\end{figure}

The particle-exchange framework is a toy model which can never capture the underlying physics described by quantum field theories. In terms of predicting the correct functional form of long-range forces, the model appears to have two intrinsic drawbacks which prevent it from describing familiar forces: (1) fields are continuous and do not carry a lag time proportional to $r\,$ between infinitesimal impulses, (2) the particle-exchange model is intrinsically one dimensional, whereas the fields giving rise to electromagnetism and gravity spread out in (at least) three dimensions. We can imagine rectifying the first drawback by promoting the mediating particles to a sort of continuous, incompressible fluid which continuously delivers impulse to the massive particles. Taking $\Delta t\,$ to be a short interval of time and assuming each differential volume element of fluid is reflected as before, the impulse delivered to the massive particle is
\begin{equation}
\Delta p = 2\rho A c^{2}\Delta t,
\end{equation}
where $\rho\,$ is fluid mass density and $A\,$ the effective cross sectional area of the particle. This results in a {\it constant} effective force,
\begin{equation}
F' \rightarrow \frac{\Delta p}{\Delta t} = 2\rho A c^{2}.\label{eq:force3}
\end{equation}
While this appears to make matters worse, it actually results in a more accurate reflection of the effective dimensionality. To explore this point, let us consider a simple example of the role of dimensionality provided by electrostatics. Gauss' law for the electric field due to a particular charge density configuration $\rho\,$ states~\cite{GriffithsEM}
\begin{equation}
\int {\bf E}\cdot d{\bf A} = \int\frac{\rho}{\epsilon_{0}}d\tau.
\end{equation}
For point-like sources, the density is sharply peaked around a single point in space $\rho \sim \delta^{3}({\bf r}- {\bf r}_{0})$, and the field spreads out into all three dimensions. For an infinite line of charge (say along the $z$-axis), we could imagine stacking infinitesimal units $dq = \lambda dz\,$ on top of each other for some finite linear charge density $\lambda$. By stacking charges like this, we cancel the $z$-component of the differential field $d{\bf E}\,$ and enhance the component in the $xy$-plane. The field of an infinite wire decays more slowly ($E\sim r^{-1}$) than for a point charge ($E\sim r^{-2}$), and we may think of this as a reduction in the dimensions accessible to the field. By occupying an entire linear dimension with charge density, we only have two remaining dimensions for the field to access, so the field is effectively strengthened. The situation is even more extreme when charge density spans two dimensions, as for an infinite sheet of charge. In this case only one dimension remains in which the field may distribute itself, and the result is a constant field which does not decay at all. All three configurations are sketched in figure~\ref{fig:efields}.
A test charge brought into any of the fields produced by the three configurations in figure~\ref{fig:efields} would feel a force $F = qE$. Depending on the effective dimensionality of the field, this force will obey a different power law according to
\begin{equation}
F \propto \left\{\begin{array}{cc} r^{-2} & (d_{\mbox{\scriptsize eff}} = 3)\\
r^{-1} & (d_{\mbox{\scriptsize eff}} = 2)\\
r^{0} & (d_{\mbox{\scriptsize eff}} = 1).\end{array}\right.\label{eq:fdim}
\end{equation}
An example of a naturally occurring system with an effectively ``one-dimensional'' field configuration is that of mesons and hadrons, which are bound states of two or three quarks, respectively. As the quarks move apart, the gluon field is compressed into one-dimensional ``flux tubes,'' containing all the field energy along a narrow tube and resulting in a confinement force which is constant as separation distance increases~\cite{PeskinSchroeder}. Equivalently, the potential energy grows linearly with separation distance. Though the quantum chromodynamics involved is quite complex and gives rise to an attractive force between bound quarks (or a quark and antiquark), the basic field geometry can be understood by this analogy with electrostatics.

Thus, the prediction of~(\ref{eq:force3}) that a continuous fluid results in a constant force is actually quite appropriate for the effectively one-dimensional problem under consideration. This conclusion is a bit of a sleight-of-hand, because for such a fluid to exist, the massive particles would have to continuously ``emit'' mass in such a way that keeping track of the increasing amount of fluid being ``reflected'' would appear to enhance the interaction into a runaway cascade of momentum transfer. However, for a fluid that is able to consistently spread out in three dimensions, the two additional inverse powers of $r$ from~(\ref{eq:fdim}) do transform the basic form in~(\ref{eq:force}) into a $r^{-2}$ Coulomb-like interaction. 

The main result in this section is that the particular power law we obtain results from a combination of collision rate and effective dimensionality. Using basic dimensional analysis, it is possible predict how modifications to the model result in different emergent force laws. To adjust the model in such a way that an inverse-square law is obtained from physically sound assumptions, one requires a sort of ``fluid-like'' entity which transmits momentum, and presumably energy. The entity which fits this description {\it and} can be defined in way that respects special relativity is a classical field, such as the electromagnetic field. Elementary considerations about energy in interacting systems suggest the necessity of fields when considering action-at-a-distance interactions~\cite{Chabay}. The analysis in this section provides an alternative motivation for introducing the role of fields as mediators of effective interactions. 

\section{Summary}\label{sec:summary}
In this paper we thoroughly examined a simple model for classical interactions through the exchange of mediating particles in which momentum conservation is enforced for each collision and the mediating particle speed is fixed at a constant value, $c$. As demonstrated in simulations and analytic reasoning, the resulting interactions yield an effectively conservative theory at low energies with a $1/r$ force. The ``non-relativistic'' approximation breaks down at high energies. Regardless of initial separation, the massive particles both eventually reach the maximum speed allowed by the physical mechanism of energy transfer within the system. Notably, any actual static potential such as Coulomb's law also breaks down when the relative speeds of interacting charges approach the speed of light. This relativistic breakdown of the notion of static potential energy functions is rarely treated at the introductory level.

The classical particle exchange analogy of ice skaters throwing a ball back and forth has typically been used as an illustration in public outreach presentations and in teaching, from general education science courses to introductory and advanced physics courses. However, the analogy has value as a physical system for students to investigate quantitatively. The phenomenon presented can be used in various contexts including homework, an in-class activity, a computational physics exercise, or assessment. Furthermore, it can be used at both the introductory and advanced level in the undergraduate curriculum. 

Investigating the emergence of force laws from classical particle in the toy model considered can serve as a valuable experience for students. A thorough examination of this particle-exchange model requires students to come to terms with the assumptions, limitations, improvements and approximations while also making comparison of numerical and analytic predictions. Such a process encapsulates the myriad of ways model building and examination is performed in theoretical physics. By employing the notion of emergence in physical theories as an motivating topic, students gain experience in the process of model building and investigation as employed by researchers in theoretical physics~\cite{Andaloro1991}. While classical particle exchange is far too simplified from reality to deliver quantitatively accurate predictions for the nature of emergent force laws, it is useful as a simple setting for exploring the {\it mechanism} of emergence in physical theories. This use of toy models to investigate mechanisms is similar to how unrealistic quantum field theories such as $\phi^{3}$ theory in six dimensions can be used to explore complex procedures such as renormalization in a simplified setting~\cite{Srednicki}.

In introductory physics, students learning computational modeling~\cite{Chabay} can investigate the phenomenon numerically. Derivation of the change in speed of a massive particle, $\delta v = 2mc/M\,$, using conservation of momentum (see (\ref{Eq:momentumconservation})) is a straightforward exercise in introductory physics. Students can also explore and describe the position-time and velocity-time graphs. Because position and velocity change abruptly, introductory students have the opportunity to fit a smooth function to values that change discretely.  Furthermore, teachers can use this system to assess understanding of potential energy functions such as (\ref{eq:potentialenergy}) and conservation of energy. Having already studied systems of particles interacting via the inverse-square law, students can practice applying a similar analysis to the $1/r$ force, possibly preparing them for similar forces that arise in an E\&M course. Finally, as shown in this paper, teachers can also use the system as an application in a junior/senior level course in mechanics~\cite{Timberlake} or mathematical physics where students are expected to explore the limits of the model using more advanced computational and analytical techniques.\\[7ex]

%\begin{acknowledgments}
%\end{acknowledgments}


\begin{thebibliography}{15}

\bibitem{BauerWestfall} Bauer W and Westfall GD 2011 {\it University Physics with Modern Physics}, 1st ed. (New York: McGraw-Hill)

\bibitem{Mazur} Mazur E 2015 {\it Principles and Practice of Physics} (Boston: Pearson)

\bibitem{Giancoli} Giancoli DC 2000  {\it Physics for Scientists and Engineers with Modern Physics}, 3rd ed. (Upper Saddle River, NJ: Prentice Hall)

\bibitem{YoungFreedman} Young HD and Freedman RA 2008 {\it Sears and Zemansky's University Physics}, 12th ed. (San Francisco: Pearson)

\bibitem{Hilborn} Hilborn RC 2014 What should be the role of field energy in introductory physics courses? {\it Am. J. Phys.} {\bf 82}(1) 66--71

\bibitem{Etkina} Etkina E, Warren A and Gentile M 2006 The Role of Models in Physics Instruction {\it Phys. Teacher} {\bf 44}(1) 34--39

\bibitem{Zee} Zee A 2010 {\it Quantum field theory in a nutshell}, 2nd ed. (Princeton, NJ: Princeton U Press)

\bibitem{Harney} Harney RC 1973 A Method for Obtaining Force Law Information by Applying the Heisenberg Uncertainty Principle {\it Am. J. Phys.} {\bf 41}(1) 67--70

\bibitem{GriffithsPart} Griffiths DJ 1987 {\it Introduction to Elementary Particles} (Hoboken, NJ: Wiley)

\bibitem{Dunne} Dunne P 2002 A reappraisal of the mechanism of pion exchange and its implications for the teaching of particle {\it Phys. Ed.} {\bf 37} 211--222

\bibitem{Rubakov} Rubakov V 2002 {\it Classical Theory of Gauge Fields} (Princeton, NJ: Princeton U Press)

%\bibitem{note1} The physics is unchanged if we interpret this lack of interaction as a perfectly elastic collision between the mediating particles. In fact, in one dimension energy conservation and momentum conservation only allow for exchange of incoming momenta in two-particle scattering events.\cite{Sutherland}

%\bibitem{note1a} We use the term ``massive'' to describe the particles of mass $M \gg m$, but this convenient terminology is not meant to suggest that the mediating particles are massless. While we take $m \ll M$, the model {\it requires} a nonzero mediating particle mass $m$.

%\bibitem{Sutherland} Sutherland B 2004 {\it Beautiful Models: 70 Years of Exactly Solved Quantum Many Body Problems} (Hackensack, NJ: World Scientific)

%\bibitem{note2} Note that the ultimate fate of the massive particles is determined by the mass ratio $\frac{2m}{M}\,$ and initial conditions. For $v_{0}=0$, if $M\,$ divides $2m\,$ exactly, then the massive particles will ultimately be accelerated to exactly $v=c$. If $M\,$ does not exactly divide $2m$, then the final collisions will propel each mass to a speed larger than $c$, but with a difference bounded by $v_{\infty} - c < \frac{2mc}{M}$.

\bibitem{Abramowitz} Abramowitz M and Stegun IA (eds.) 1965 {\it Handbook of Mathematical Functions} pp. 877 (New York: Dover)

\bibitem{Hecht} Hecht E 2019 Understanding energy as a subtle concept: A model for teaching and learning energy {\it Am. J. Phys.} {\bf 87}(7) 495--503

\bibitem{ArfkenWeber} Arfken GB and Weber HJ 2012 {\it Mathematical Methods for Physicists}, 7th ed. (Waltham, MA: Academic Press)

\bibitem{PeskinSchroeder} Peskin ME and Schroeder DV 1995 {\it An Introduction to Quantum Field Theory} (New York: Perseus)

\bibitem{GriffithsEM} Griffiths DJ 1999 {\it Introduction to Electrodynamics}, 3rd ed. (Hoboken, NJ: Wiley)

%\bibitem{Corless} Corless RM, Gonnet GH, Hare, DEG, Jeffrey DJ and Knuth DE 1996 On the Lambert W function {\it Adv. Comp. Math.} {\bf 5} 329-359

%\bibitem{Feynman1} Feynman RP 1998 {\it The Theory of Fundamental Processes}, (Boulder, C.O.: Westview Press)

%\bibitem{FeynmanGrav} Feynman RP, Morinigo FB and Wagner WG 2002 {\it Feynman Lectures on Gravitation}, (Boulder, CO: Westview Press)

\bibitem{Ee} Ee JH and Lee J 2012 A unique pure mechanical system revealing dipole repulsion {\it Am. J. Phys.} {\bf 80}(12) 1078--1084

%\bibitem{Sinai} Sinai YG 1999 {\it Theor. Math. Phys.} {\bf 121} 1351

%\bibitem{De} De S 1997 {\it Phys. and Tech. Quest} {\bf 2}, 35

%\bibitem{Tucker} Tucker A 2012 {\it Applied Combinatorics}, $6^{\mbox{\scriptsize th}}\,$ed. (Hoboken, NJ: Wiley)

\bibitem{Chabay} Chabay R, Sherwood B and Titus A 2019 A unified, contemporary approach to teaching energy in introductory physics {\it Am. J. Phys.} {\bf 87}(7) 504--509

\bibitem{Andaloro1991} Andaloro G, Donzelli V and Sperandeo-Mineo RM 1991 Modelling in physics teaching: the role of computer simulation {\it Int. J. Sci Ed.} {\bf 13}(3), 243--254

\bibitem{Srednicki} Srednick M 2007 {\it Quanutm Field Theory}, (Cambridge, UK: Cambridge U Press)

\bibitem{Chabay} Chabay R and Sherwood B 2008 Computational physics in the introductory calculus-based course {\it Am. J. Phys.} {\bf 76} 307--313

\bibitem{Timberlake} Timberlake T and Hasbun J 2008 Computation in classical mechanics {\it Am. J. Phys.} {\bf 76} 334--339

\end{thebibliography}
\end{document}